\documentclass{easychair}
\usepackage[utf8]{inputenc}

\usepackage{float}

\usepackage{doc}
\usepackage{graphicx}
\usepackage{syntax}

\usepackage{booktabs}

\usepackage{subcaption}

\usepackage{multicol}
\usepackage{tikz}
\usetikzlibrary{shapes,arrows,positioning,fit,calc,patterns,tikzmark}

\usepackage{listings}
\lstdefinestyle{mystyle}{
    basicstyle=\ttfamily\footnotesize,
}
 
\usepackage{xspace}
\newcommand{\tezos}{\textsc{Tezos}\xspace}
\newcommand{\tezla}{\textsc{Tezla}\xspace}
\newcommand{\softc}{\textsc{SoftCheck}\xspace}
\newcommand{\scil}{\textsc{SCIL}\xspace}
\newcommand{\mich}{\textsc{Michelson}\xspace}

\title{\tezla, an intermediate representation for static analysis of \mich
    smart contracts\thanks{This work was supported and funded by the Tezos
        Foundation by the project FRESCO (Formal Verification of Tezos Smart
        Contracts).}}
\titlerunning{Tezla, an intermediate representation for \mich}
\author{João Santos Reis\inst{1} \and Paul Crocker\inst{2} \and Simão Melo de Sousa\inst{1,2}}
\authorrunning{J. Reis and P. Crocker and S. Melo de Sousa}
\institute{Nova-Lincs, University of Beira Interior, Portugal \\
    \email{joao.reis@ubi.pt}
    \and
   Release Lab, C4 and  University of Beira Interior, Portugal\\
    \email{\{desousa|crocker\}@di.ubi.pt}}

\begin{document}
\maketitle              
\begin{abstract}
    This paper introduces \tezla, an
    intermediate representation of \mich smart contracts that eases the design of static smart contract analysers. This intermediate representation uses a store and preserves the semantics, flow and resource usage of the original smart
    contract. This enables properties like gas consumption to be statically verified. We provide an automated decompiler of \mich~smart contracts to \tezla.
    In order to support our claim about the adequacy of \tezla, we develop a static analyser that takes advantage of the \tezla~representation of \mich~smart contracts to prove simple but non-trivial properties.


\end{abstract}
\section{Introduction}
The term ``smart contract'' was proposed by Nick Szabo as a way to formalize and
secure relationships over public networks~\cite{Szabo1997}. In a blockchain,
a smart contract is an application written in some specific language that is
embedded in a transaction (hence the program code is immutable once it is out in
the network). Some examples of smart contracts applications are the management
of agreements between parties without resorting to a third party (escrow) and to
function as a multi-signature account spending requirement. Smart contracts have
the ability to transfer/receive funds to/from users or from other smart contracts and
can interact with other smart contracts.

There has been recent reports of bugs and consequently attacks in smart
contracts that have led to losses of millions of dollars worth of assets. One of the
most famous and most costly of these attacks was on the Distributed Autonomous
Organization (DAO), on the Ethereum blockchain. The attacker managed to withdraw
approximately 3.6 million ether from the contract.

Given the fact that a smart contract in a blockchain can't be updated or patched,
there is an increasing interest in providing tools and mechanisms that guarantee or
potentiate the correctness of smart contracts and to verify certain properties.

However, current tools and algorithms for program verification, based for example on deductive verification and static analysis, are usually designed for classical store-based
languages in contrast with \mich, the smart contract language for the Tezos Blockchain~\cite{Goodman2014,michelson}, which is stack based.

To facilitate the usage of such tools to verify \mich~smart
contracts, we present \tezla, a store based intermediate  representation language for \mich, and its respective tooling.
We provide an automated decompiler of \mich~smart contracts to \tezla.
The decompiler  preserves the semantics, flow and resource usage of the original smart contract, so that
properties like gas consumption can be faithfully verified at the \tezla~representation level.
To support our work, we present a case-study of a demo platform for the static analysis of Tezos smart contracts using the \tezla~intermediate representation alongside with an example analysis.

The paper is structured as follows. In section 2 we introduce the syntax and
semantics of \tezla. The decompiler mechanism is described in section 3. Section
4 addresses the static analysis platform case-study that targets
\tezla-represented smart contracts. Finally section 5 concludes with a general
overview of this contribution and future lines of work.



\section{Tezla}

\tezla aims to facilitate the adoption of existing static analysis tools and algorithms.
As such, \tezla is an intermediate representation of \mich code that uses a store instead of a stack, enforces the Static Assignment Form (SSA) and preserves information about gas consumption. We will see in the next section how such characteristics ease the translation of \tezla program into their Control Flow Graph (CFG) forms and the construction of data-flow equations.

Compiled languages (like Albert, LIGO, SmartPy, Lorentz, etc.) also provide a higher-level abstraction over Michelson. However, as it  happens with most compiled languages, the produced code may not be as concise or compact as expected which, in the case of smart contracts, may result in undesired costs. \tezla was designed to have a tight integration with the Michelson code to be executed, not as a language that compiles to it nor a higher level language that ease the writing of \mich smart contracts.

In the \tezla representation, push-like instructions are translated into variable assignments, whereas instructions that consume stack values are transformed to expressions that use as arguments the variables that match the values from the stack. Furthermore, lists, sets and maps deconstruct and lifting of \texttt{option} and \texttt{or} types that happen implicitly are represented through explicit expressions added to \tezla.

Since the operational effect of stack manipulation is transposed into variable assignments, we also expose in a \tezla represented contract the stack manipulation as instructions that act as no-op instructions in the case of a semantics that do not take resource consumption into account\footnote{This is the case of the semantics presented in this paper.}. In the case of a resource aware semantics, these instructions will semantically encode this consumption.

The following section describes in detail the process of transforming a \mich smart contract to a \tezla representation.

\subsection{Push-like instructions and stack values consumption}

Instructions that push \(N\) values to the stack are translated to \(N\) variable
assignments of those values. The translation process maintains a \mich program stack that associates each stack position to the variable to which that
position value was assigned to. When a stack element is consumed, the
corresponding variable is used to represent the value. A very simple example is provided in fig.~\ref{fig:example1}.
\begin{figure}[ht]
    \begin{subfigure}[b]{0.4\textwidth}
    \begin{lstlisting}
PUSH nat 5;
PUSH nat 6;
ADD;
    \end{lstlisting}
    \caption{\mich code.}%
    \label{fig:example1mich}
    \end{subfigure}
    \hfill
    \begin{subfigure}[b]{0.4\textwidth}
    \begin{lstlisting}
v1 := PUSH nat 5;
v2 := PUSH nat 6;
v3 := ADD v1 v2;
\end{lstlisting}
    \caption{\tezla code.}%
    \label{fig:example1tezla}
    \end{subfigure}
    \caption{Stack manipulation example.}%
    \label{fig:example1}
\end{figure}

The block on figure~\ref{fig:example1mich} is translated to the \tezla~representation shown in figure~\ref{fig:example1tezla}.

From the previous example, we can also observe that \mich~instructions that
consume \(N\) stack variables are translated to an expression that consumes those
\(N\) values. Concretely, the instruction \texttt{ADD} that consumes two values
(say, \texttt{a} and \texttt{b}), from the stack is translated to \texttt{ADD a b}.

\subsection{Branching}

\mich~provides developers with branching structures that act on different conditions. As \tezla~aims at being used as an intermediate representation for static analysis, there are some properties we would like to maintain. One such property is static single assignment form (SSA-form)~\cite{Rosen1988}.
This is guaranteed as \tezla-represented smart contracts are, by construction, in SSA-form, since each assignment uses new variables.

In order to deal with branching, the \tezla~representation makes use of \(\phi \)-functions (see~\cite{Rosen1988}) that select between two values depending on the branch.
As an illustration consider the \mich~example in figure~\ref{fig:exampleTz1Mich}.

\begin{figure}[ht]
    \centering
    \begin{subfigure}[t]{0.47\textwidth}
        \centering
        \begin{lstlisting}
parameter int ;
storage (list int) ;
code { UNPAIR ;
       SWAP ;
       IF_CONS
         { DUP ; DIP { CONS ; SWAP } ;
           ADD ; CONS }
         { NIL int ; SWAP ; CONS } ;
       NIL operation ;
       PAIR }
        \end{lstlisting}
        \caption{\mich code.}%
        \label{fig:exampleTz1Mich}
    \end{subfigure}
    \hfill
    \begin{subfigure}[t]{0.4\textwidth}
        \centering
        \begin{lstlisting}[escapeinside={(*}{*)}]
v0 := CAR parameter_storage;
v1 := CDR parameter_storage;
SWAP;
IF_CONS v1
{
  v2 := hd v1;
  v3 := tl v1;
  v4 := DUP v2;
  v5 := CONS v2 v3;
  SWAP;
  v6 := ADD v4 v0;
  v7 := CONS v6 v5
}
{
  v8 := NIL int;
  SWAP;
  v9 := CONS v0 v8
};
v10 := (*\(\phi\)*)(v7, v9);
v11 := NIL operation;
v12 := PAIR v11 v10;
        \end{lstlisting}
        \caption{\tezla code.}%
        \label{fig:exampleTz1Tezla}
    \end{subfigure}
    \caption{Branching example.}%
    \label{fig:exampleTz1}
\end{figure}

This contract takes an \texttt{int} as parameter and a list of \texttt{int}s
as storage and inserts the sum of the parameter with the head of the list at the
lists's head. If the list is empty, it inserts the parameter into the empty list.
Here, each branch of the \texttt{IF_CONS} instruction will result in a stack
with a list of integers, whose values depends on which branch was executed.

This translates to the \tezla~representation presented figure~\ref{fig:exampleTz1Tezla}.

The variable \texttt{v10} will receive its value through a \(\phi \)-function that
returns the value of \texttt{v7} if the true branch is executed, or the value of
\texttt{v9} otherwise.

The \texttt{IF_CONS} instruction deconstructs a list in the true branch, putting
the head and the tail of the list on top of the stack. From this example, it is
possible to observe that the deconstruction of a list is explicit through two
variable assignments. This is also the behaviour of \texttt{IF_NONE} and
\texttt{IF_LEFT} instructions, where the unlifting of \texttt{option} and
\texttt{or} types happens explicitly through an assignment.

\subsection{Loops, maps and iterations}
\mich~also provides language constructs for looping and iteration over the
elements of lists, sets and maps. These are treated using the same
\(\phi \)-functions mechanism in order to preserve SSA-form. We can observe this
on the example fig.~\ref{fig:exampleTz2}.

\begin{figure}[ht]
    \centering
    \begin{subfigure}[t]{0.5\textwidth}
        \centering
        \begin{lstlisting}[escapeinside={(*}{*)}]
PUSH nat 0 ;
LEFT nat ;
LOOP_LEFT
 { DUP ;
   PUSH nat 100 ;
   COMPARE ;
   GE ;
   IF
    { PUSH nat 1 ;
      ADD ; LEFT nat }
    { RIGHT nat } } ;
INT ; 
        \end{lstlisting}
        \caption{\mich code.}%
        \label{fig:exampleTz2Mich}
    \end{subfigure}
    \hfill
    \begin{subfigure}[t]{0.49\textwidth}
        \centering
        \begin{lstlisting}[escapeinside={(*}{*)}]
v0 := PUSH nat 0;
v1 := LEFT nat v0;
LOOP_LEFT v2 := (*\(\phi\)*)(v1, v12)
{
  v3 := unlift_or v2;
  v4 := DUP v3;
  v5 := PUSH nat 100;
  v6 := COMPARE v5 v4;
  v7 := GE v6;
  IF v7
  {
    v8 := PUSH nat 1;
    v9 := ADD v8 v3;
    v10 := LEFT nat v9;
  }
  {
    v11 := RIGHT nat v3;
  }
  v12 := (*\(\phi\)*)(v10, v11);
}
v13 := unlift_or v2;
v14 := INT v13;
        \end{lstlisting}
        \caption{\tezla code.}%
        \label{fig:exampleTz2Tezla}
    \end{subfigure}
    \caption{Loop example.}%
    \label{fig:exampleTz2}
\end{figure}

This example uses a \texttt{LOOP_LEFT} (loop with an accumulator) to sum 1
to a \texttt{nat} (starting with the value 0) until that value becomes greater
than 100 and casts the result to an \texttt{int}.
This example translates to the code presented in fig~\ref{fig:exampleTz2Tezla}.

Note that the \texttt{LOOP_LEFT} variable is assigned to the value of
\texttt{v1} if it is the first time that the loop condition is checked, or
\texttt{v12} if the program flow comes from the loop body. Also notice that the
same explicit deconstruction of an \texttt{or} variable is applied here, where
\texttt{v5} gets assigned the value of the unlifting of the loop variable in the
beginning of the loop body and at the end of the loop. Similar behaviour applies
to the other looping and iteration instructions.

\subsection{Full example}
We now present a full example of a complete \mich~smart contract (figure~\ref{fig:contractexampleMich}).

\begin{figure}[ht]
    \centering
    \begin{subfigure}[t]{0.47\textwidth}     \begin{lstlisting}[escapeinside={(*}{*)}]
parameter (list bool) ;
storage (pair bool (pair nat int)) ;
code { DUP ;
       CAR ;
       DIP { CDR } ;
       DIP { DUP ; CAR ; DIP { CDR } ;
             DIP { DUP ; CAR ;
                   DIP { CDR } } } ;
       ITER { AND ;
              DUP ;
              IF
               { DIP 2
                     { PUSH int 1 ;
                       ADD } }
               { DIP 2
                     { PUSH int -1 ;
                       ADD } } } ;
       DIP { PAIR } ;
       PAIR ;
       NIL operation ;
       PAIR }
        \end{lstlisting}
        \caption{\mich code.}%
        \label{fig:contractexampleMich}
    \end{subfigure}
    \hfill
    \begin{subfigure}[t]{0.45\textwidth}
            \begin{lstlisting}[escapeinside={(*}{*)}]
v0 := DUP parameter_storage;
v1 := CAR v0;
v2 := CDR parameter_storage;
v3 := DUP v2;
v4 := CAR v3;
v5 := CDR v2;
v6 := DUP v5;
v7 := CAR v6;
v8 := CDR v5;
ITER v9 := (*\(\phi\)*)(v1, v18)
{
    v10 := hd v9;
    v11 := AND v10 v4;
    v12 := DUP v11;
    IF v12
    {
        v13 := PUSH int 1;
        v14 := ADD v13 v8;
    }
    {
        v15 := PUSH int -1;
        v16 := ADD v15 v8;
    };
    v17 := (*\(\phi\)*)(v14, v16);
    v18 := tl v9;
};
v19 := PAIR v7 v17;
v20 := PAIR v11 v19;
v21 := NIL operation;
v22 := PAIR v21 v20;
return v22;
    \end{lstlisting}
        \caption{\tezla code.}%
        \label{fig:contractexampleTezla}
    \end{subfigure}
    \caption{Example contract.}%
    \label{fig:contractexample}
\end{figure}

The contract takes a list of \texttt{bool}s as parameter and iterates over that list, It performs a boolean \texttt{AND} between an element of the list and the previous \texttt{AND} (the initial value of this accumulator is the \texttt{bool} on the storage). Depending on the result it either adds 1 ot -1 to the \texttt{int} on the storage. The values to be stored are the last \texttt{AND} result, the \texttt{nat} that was previously on the storage (notice that this value isn't changed nor it is used anywhere else in the program) and the resulting \texttt{int} from the sums on the iteration. This contract translates to the \tezla~code of fig.~\ref{fig:contractexampleTezla}.

In this complete example we can observe that a \mich contract has a parameter and storage.
The initial stack of any \mich~smart contract is a stack that contains a single pair whose first element is the input parameter and second element is the contract storage. As such, we introduce a variable called \texttt{parameter_storage} that contains the value of that pair.

The final stack of any \mich~smart contract is also a stack that contains a single pair whose first element is a list of internal operations that it wants to emit and whose second element is the resulting storage of the smart contract.
We identify the variable containing this pair through the addition of a \texttt{return} instruction.





\section{Building statics analyses for Tezla smart contracts}

In this section, we present the experiments conducted in order to test and
demonstrate the applicability of the \tezla intermediate representation to
perform static analysis.

\subsection{SoftCheck}

We build and organise these static analyses upon a generic data-flow analysis platform
called \softc~{\cite{reisSoftcheck}}.
\softc provides an internal and intermediate program  representation, called \scil, rich enough to express high-level as well as low-level imperative programming constructs and simple enough to be adequately translated into CFGs.

\softc is organised upon a generic monotone
framework~\cite{Kam1977} that is able to extract a set of data-flow equations from (1) a suitable representation of programs and; (2) a set of monotone functions; and then to solve them. \softc is written in \textsc{OCaml} and makes use of functor interfaces to leverage its genericity (see fig~\ref{fig:softchek}).

By generic we mean that, given a translation from a programming language to \scil.
\softc gives the ability to instantiate its underlying monotone framework by means of a functor interface. Then all defined static analyses are automatically available for the given programming language.

On the other hand, once written as a set of properties and monotone functions,  a particular static analysis can be incorporated  (again, through  instantiating a functor) as an available static analysis for all interfaced programming languages.

\softc offers several standard data-flow analysis such as very busy expressions, available expressions, tainted analysis etc.

We propose in the next sections to detail how we have interfaced \tezla with \scil, how we have designed a simple but useful data-flow analysis within \softc  and how we have tested this analysis on the \mich smart contracts running in the \tezos blockchain.





\begin{figure}[ht]
    \centering
	\resizebox{0.7\textwidth}{!}{\pgfdeclarelayer{background}
    \pgfdeclarelayer{foreground}
    \pgfsetlayers{background,foreground}
    \begin{tikzpicture}[
            node distance=7mm,
            typetag/.style={rectangle, draw, font=\scriptsize\ttfamily, anchor=west},
        ]
        \begin{pgfonlayer}{foreground}
            \node (l1) {\texttt{Language1}};
            \node (ast1) [below=of l1.west, typetag] { Ast };
            \node (cfg1) [right=of ast1.east, typetag] { Cfg };
            \node (la11) [below=of ast1.west, typetag] { Analysis1Language1 };
            \node (la12) [below=of la11.west, typetag] { Analysis1Language2 };
            \node (l1dots) [below=of la12.north] { $\vdots$ };
            \node (ll1) [draw, fit={(l1) (ast1) (cfg1) (la11) (la12) (l1dots)}] {};

            \node (l2) [below=of l1.south, yshift=-3.6cm] {\texttt{Language2}};
            \node (ast2) [below=of l2.west, typetag] { Ast };
            \node (cfg2) [right=of ast2.east, typetag] { Cfg };
            \node (la21) [below=of ast2.west, typetag] { Analysis1Language2 };
            \node (la22) [below=of la21.west, typetag] { Analysis1Language2 };
            \node (l2dots) [below=of la22.north] { $\vdots$ };
            \node (ll2) [draw, fit={(l2) (ast2) (cfg2) (la21) (la22) (l2dots)}] {};

            \node (ldots) [below=of ll2.south] {$\vdots$};

            \node (label1) [below=of ldots.south, align=center] {Language specific \\ input};


            \node (a1) [right=of l1.east, xshift=+1.5cm] {\texttt{Analysis1}};
            \node (p1) [below=of a1.west, typetag] { Properties };
            \node (f1) [right=of p1.east, typetag] { Monotone functions };
            \node (aa1) [draw, fit={(a1) (p1) (f1)}] {};

            \node (a2) [below=of a1.south, yshift=-1.6cm] {\texttt{Analysis2}};
            \node (p2) [below=of a2.west, typetag] { Properties };
            \node (f2) [right=of p2.east, typetag] { Monotone functions };
            \node (aa2) [draw, fit={(a2) (p2) (f2)}] {};

            \node (adots) [below=of aa2.south] { $\vdots$ };

            \node (label2) [below=of adots.south, align=center] {Analysis specific \\ input};


            \node (lattices) [right=of a1.east,draw,xshift=+5.1cm] {\texttt{Lattices}};
            \node (dependencies) [below=of lattices,draw] {\texttt{Dependences}};
            \node (fix) [draw,below=of dependencies,xshift=-1.66cm] {\texttt{Fix}};
            \node (cfggen) [right=of fix.east,draw] {\texttt{CfgGenerator}};
            \node (hidden) [below=of dependencies] {};
            \node (label4) [below=of hidden] {Support libraries};


            \node (framework) [draw,below=of label4,yshift=-1.0cm] {\texttt{Framework}};
            \node (solver) [draw,below=of framework.south,yshift=-2cm] {\texttt{Solver}};
            \node (program) [right=of framework.east,xshift=+1.2cm,draw,rounded corners,label=Program] {};
            \node (result) [draw,right=of solver.east,rounded corners,label=Result,fill=black,xshift=+1.5cm] {};
            \node (label3) [below=of solver.south, align=center] {Solver engine};


        \end{pgfonlayer}
        \begin{pgfonlayer}{background}
            \node (languages) [rounded corners,fill=gray!15,draw,fit={(ll1) (ll2) (label1)}] {};
            \node (analyses) [rounded corners,fill=yellow!15,draw,fit={(aa1) (a2) (label2)}] {};
            \node (engine) [rounded corners,fill=green!15,draw,fit={(framework) (solver) (label3)}] {};
            \node (libs) [rounded corners,fill=green!15,draw,fit={(lattices) (dependencies) (fix) (cfggen) (label4)}] {};
        \end{pgfonlayer}
        \begin{pgfonlayer}{foreground}
            \path[->,>=angle 90]
            (ll1) edge (aa1)
            (ll1) edge (aa2)
            (ll2) edge [bend left=15] (aa1)
            (ll2) edge (aa2)
            (framework) edge [bend left] (solver)
            (solver) edge [bend left] (framework)
            (solver) edge (result)
            (analyses) edge (engine)
            (program) edge (framework)
            (libs) edge (analyses)
            (libs) edge (engine);
        \end{pgfonlayer}
    \end{tikzpicture}}
    \caption{\softc~in a picture}%
    \label{fig:softchek}
\end{figure}

\subsection{Constructing a Tezla Representation of a Contract}

To obtain the \tezla~representation of a smart contract, we first developed a
parser to obtain an abstract syntax representation of a \mich smart contract.
This parser was implemented in OCaml and Menhir and respects the syntax
described in the Tezos documentation~\cite {michelson}.
It allows us to obtain a data type that fully abstracts the syntax (with the exception of annotations). To improve the integration between these two forms, \tezla data types were built upon the data types of \mich.



Control-flow graphs are a common representation among static analysis tools.
We provide a library for automatic extraction of such representation from any
\tezla-represented smart contract.
This library is based upon the control-flow generation template present on \softc. As such, control-flow graphs generated
with this library can be used with \softc without further work.
To instantiate the control-flow graph generation template, we simply provided the library with a module with functions that describe how control flows between each node.

\subsection{Sign detection: an example analysis}

\begin{figure}[ht]
    \centering
    \includegraphics[width=0.2\textwidth]{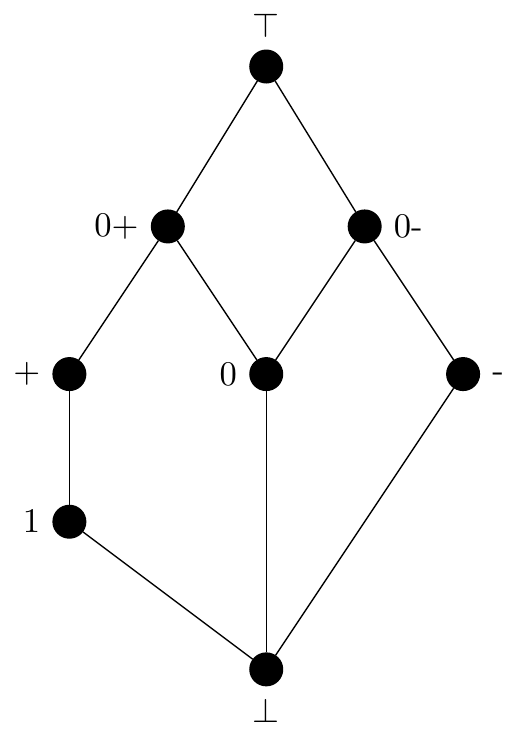}
    \caption{Sign lattice.}%
    \label{fig:signLattice}
\end{figure}

Here we devise an example of a static analysis for sign detection. The abstract domain consists of the following abstract sign values:: 0 (zero), 1 (one), 0+ (zero or positive), 0- (zero or negative), \(\top \) (don't know)
and \(\bot \) (not a number). These values are organised
according to the lattice on figure~\ref{fig:signLattice}.

Using \softc, we implemented a simple sign detection analysis of numerical values.
By definition, \texttt{nat}s have a lowest precision value
of 0+, while \texttt{int}s can have any value. Every other data type has a sign value of \(\bot \).

This implementation does not propagate information to non-simple types (\texttt{pair}, \texttt{or}, etc.), but it does perform some precision refinements on branching.


To implement such an analysis, we provided \softc, in
addition to the previously defined \tezla control-flow graph library, a module that defines how each instruction
impacts the sign value of a variable. Then, using the
integrated solver mechanism based on the monotone framework,
we are able to run this analysis on any \tezla represented smart contract.

\begin{figure}[ht]
    \centering
    \includegraphics[width=0.8\textwidth]{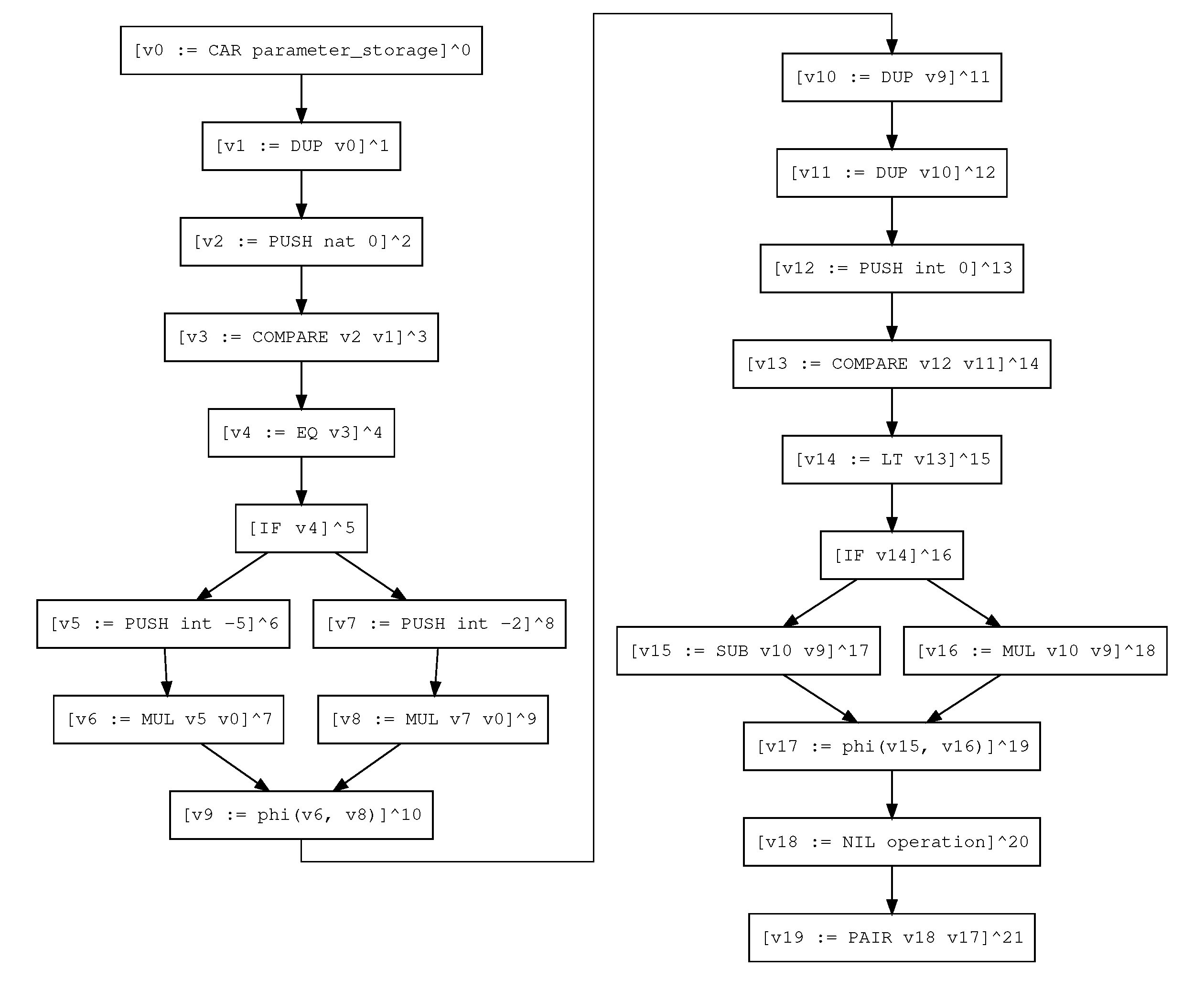}
    \caption{Generated CFG, by the \softc tool}%
    \label{fig:contractSignCfg}
\end{figure}

We now present an example. Figure~\ref{fig:contractSign} shows the code of a smart contract and its \tezla~representation. This contract
multiplies its parameter by \(-5\) if the parameter is equal to \(0\), or by
\(-2\) otherwise, and stores the result in the storage.
Figure~\ref{fig:contractSignCfg} shows the control-flow graph of
representation of that contract.

\begin{figure}[ht]
    \centering
    \begin{subfigure}[t]{0.4\textwidth}
        \begin{verbatim}
parameter nat ;
storage int ;
code { CAR ;
       DUP ;
       PUSH nat 0 ;
       COMPARE ;
       EQ ;
       IF { PUSH int -5 ; MUL }
          { PUSH int -2 ; MUL } ;
       NIL operation ;
       PAIR }
        \end{verbatim}
        \caption{\mich code.}%
        \label{fig:contractSignMich}
    \end{subfigure}
    \hfill
    \begin{subfigure}[t]{0.4\textwidth}
        \begin{verbatim}
v0 := CAR parameter_storage;
v1 := DUP v0;
v2 := PUSH nat 0;
v3 := COMPARE v2 v1;
v4 := EQ v3;
IF v4
{
    v5 := PUSH int -5;
    v6 := MUL v5 v0
}
{
    v7 := PUSH int -2;
    v8 := MUL v7 v0
};
v9 := phi(v6, v8);
v10 := NIL operation;
    \end{verbatim}
        \caption{\tezla code.}%
        \label{fig:contractSignTezla}
    \end{subfigure}
    \caption{Example contract for sign analysis.}%
    \label{fig:contractSign}
\end{figure}

Running this analysis on the previously mentioned contract
produced the results available in Figure~\ref{fig:contractSignAnalysis}.
In these results we can observe the known sign value of each variable at the exit of each block of the control-flow graph in Figure~\ref{fig:contractSignCfg}.
For brevity purposes, we omitted non-numerical variables from the result.

\begin{figure}[ht]
    \centering
    \setlength{\columnseprule}{0.25pt}
    \begin{multicols}{5}
        \begin{lstlisting}[basicstyle=\tiny]
0: {
    v0: 0+
}
1: {
    v0: 0+,
    v1: 0+
}
2: {
    v0: 0+,
    v1: 0+,
    v2: 0
}
3: {
    v0: 0+,
    v1: 0+,
    v2: 0,
    v3: 0-
}
4: {
    v0: 0+,
    v1: 0+,
    v2: 0,
    v3: 0-
}
5: {
    v0: 0+,
    v1: 0+,
    v2: 0,
    v3: 0-
}
6: {
    v0: 0,
    v1: 0,
    v2: 0,
    v3: 0-,
    v5: -
}
7: {
    v0: 0,
    v1: 0,
    v2: 0,
    v3: 0-,
    v5: -,
    v6: 0
}
8: {
    v0: +,
    v1: +,
    v2: 0,
    v3: 0-,
    v7: -
}
9: {
    v0: +,
    v1: +,
    v2: 0,
    v3: 0-,
    v7: -,
    v8: -
    }
10: {
    v0: 0+,
    v1: 0+,
    v2: 0,
    v3: 0-,
    v5: -,
    v6: 0,
    v7: -,
    v8: -,
    v9: 0-
}
11: {
    v0: 0+,
    v1: 0+,
    v2: 0,
    v3: 0-,
    v5: -,
    v6: 0,
    v7: -,
    v8: -,
    v9: 0-,
    v10: 0-
}
12: {
    v0: 0+,
    v1: 0+,
    v2: 0,
    v3: 0-,
    v5: -,
    v6: 0,
    v7: -,
    v8: -,
    v9: 0-,
    v10: 0-,
    v11: 0-
}
13: {
    v0: 0+,
    v1: 0+,
    v2: 0,
    v3: 0-,
    v5: -,
    v6: 0,
    v7: -,
    v8: -,
    v9: 0-,
    v10: 0-,
    v11: 0-,
    v12: 0
}
14: {
    v0: 0+,
    v1: 0+,
    v2: 0,
    v3: 0-,
    v5: -,
    v6: 0,
    v7: -,
    v8: -,
    v9: 0-,
    v10: 0-,
    v11: 0-,
    v12: 0,
    v13: 0+
    }
15: {
    v0: 0+,
    v1: 0+,
    v2: 0,
    v3: 0-,
    v5: -,
    v6: 0,
    v7: -,
    v8: -,
    v9: 0-,
    v10: 0-,
    v11: 0-,
    v12: 0,
    v13: 0+
}
16: {
    v0: 0+,
    v1: 0+,
    v2: 0,
    v3: 0-,
    v5: -,
    v6: 0,
    v7: -,
    v8: -,
    v9: 0-,
    v10: 0-,
    v11: 0-,
    v12: 0,
    v13: 0+
}
17: {
    v0: 0+,
    v1: 0+,
    v2: 0,
    v3: 0-,
    v5: -,
    v6: 0,
    v7: -,
    v8: -,
    v9: +,
    v10: +,
    v11: +,
    v12: 0,
    v13: 0+,
    v15: top
}
18: {
    v0: 0+,
    v1: 0+,
    v2: 0,
    v3: 0-,
    v5: -,
    v6: 0,
    v7: -,
    v8: -,
    v9: 0-,
    v10: 0-,
    v11: 0-,
    v12: 0,
    v13: 0+,
    v16: 0+
}
19: {
    v0: 0+,
    v1: 0+,
    v2: 0,
    v3: 0-,
    v5: -,
    v6: 0,
    v7: -,
    v8: -,
    v9: 0-,
    v10: 0-,
    v11: 0-,
    v12: 0,
    v13: 0+,
    v15: top
    v16: 0+,
    v17: top
}
\end{lstlisting}
    \end{multicols}
    \caption{generated report for the sign analysis}%
    \label{fig:contractSignAnalysis}
\end{figure}

It it possible to observe from the results that the analysis takes
into account several details. For instance, the sign of values of type \texttt{nat} are, by definition, always zero or positive.
The analysis also refines the sign values on conditional branches according to the test. In this case, we can notice that in blocks 6 and 7 (true branch) the sign value of \texttt{v1} must be \texttt{0}, as the test corresponds to \texttt{0 == v1}. Complementary to this, in blocks 8 and 9 the value of \texttt{v1} assumes the sign value of \texttt{+}, given that being a \texttt{nat} value its value must be \texttt{0+} and we know that its values is not zero because the test \texttt{0 == v1} failed.

We can also conclude from the result of this analysis that the block 17
(true branch) will never be carried out, as the test of that conditional
(\texttt{0 < v11}) will always be false because the sign of \texttt{v11}
is \texttt{0-}, which means it will always be less than 0.

Due to the \tezla nature, we were able to take advantage of existing of tooling,
such as the \softc~platform, and effortlessly design the run a data-flow
analysis. This enables and eases the development of static analysis that
can be used to verify smart contracts but also to perform code optimisations,
such as dead code elimination. Albeit simple, the sign analysis can be used to
instrument such dead code elimination procedure.


\subsection{Experimental Results and Benchmarking}

\tezla and all the tooling are implemented in OCaml and
are available under~\cite{fresco}.
\tezla accepts Michelson contracts that are valid according
to the Tezos protocol 006 Carthage. We conducted Experimental
evaluations that consisted in transforming to \tezla and running
the developed analyses on a batch of smart contracts.

In order to so, we implemented a tool that allows the extraction
of smart contracts available in the Tezos blockchain. With that
tool, we extracted 142 unique smart contracts. We tested
these unique contracts alongside 21 smart contracts we have implemented ourselves.

We successfully converted all smart contracts with a coverage result of all Michelson instructions except for 9 instruction that were not used in any of these 163 contracts. On those, we ran the available analyses and obtained the benchmarks presented on table~\ref{tab:benchmark}. These experiments were performed on
a machine with an Intel i7--8750H (2.2 GHz) with 6 cores and 32 GB of RAM.

In the absence of an optimisation tool that takes advantages of the information computed by the analysis, the reports produced by the analysis need to be manually inspected.  These reports, the source code of contracts under evaluation, as well as the respective analysis result  and other performed static analyses are available at~\cite{tezcheck,ReisTezCheckAnalysisResultsGitLab}.

\begin{table}[ht]
    \centering
\begin{tabular}{@{}lllll@{}}
\cmidrule(r){1-2} \cmidrule(l){4-5}
Average time      & 0.48 s                                                              &  & \begin{tabular}[c]{@{}l@{}}Worst-case\\ (number of\\ instructions)\end{tabular} & \begin{tabular}[c]{@{}l@{}}2231\\ (6.08 s)\end{tabular} \\ \cmidrule(r){1-2} \cmidrule(l){4-5} 
Worst-case (time) & \begin{tabular}[c]{@{}l@{}}9.87 s\\ (926 instructions)\end{tabular} &  & \begin{tabular}[c]{@{}l@{}}Average time\\ per instrucion\end{tabular}           & 0.0009                                                  \\ \cmidrule(r){1-2} \cmidrule(l){4-5} 
\end{tabular}
    \caption{Benchmark results.}
    \label{tab:benchmark}
\end{table}



\section{Related Work}

Albert~\cite{Bernardo2020} is an intermediate language for the development of
Michelson smart contracts. This language provides an high-level abstraction of
the stack and some of the language datatypes. This language can be compiled to
Michelson through a compiler written in Coq that targets
Mi-Cho-Coq~\cite{Bernardo2019}, a Coq specification of the Michelson language.


Several high-level languages \cite{Alfour,Andrews,Maurel,DaiLambda,Serokell} that target
Michelson have been developed. Each one presents a different mechanism that
abstracts the low-level stack usage. However, a program analysis tool that
would target one of these languages should not be easily reusable to
programs written in the other languages.



Scilla~\cite{Sergey2018,sergeySaferSmartContract2019} is an intermediate
language that aims to be a translation target of high-level languages for smart
contract development. It introduces a communicating automata-based computational
model that separates the communication and programming aspects of a contract.
The purpose of this language is to serve as a basis representation for program
analysis and verification of smart contracts.

Slither~\cite{Feist2019}, presented in 2019, is a static analysis framework for
Ethereum smart contract. It uses the Solidity smart contract compiler
generated Abstract Syntax Tree to transform the contract into an intermediate
representation called SlithIR. This representation also uses a SSA form and
reduced instruction in order to facilitate the implementation of program
analyses of smart contracts. However Slither has no formal semantics and also
the representation is not able to accurately model some low level information like gas computations.

Solidifier~\cite{Antonino2020} is a bounded model checker for Ethereum smart
contracts that converts the original source code to Solid, a formalisation of
Solidity that runs on its own execution environment. Solid is translated to
Boogie, an intermediate verification language that is used by the bounded model
checker Corral, which it then used to look for semantic-property violations.

Durieux et.\ al~\cite{Durieux2020} presented a review on static analysis tools for Ethereum smart contracts. This work presents an extensive list of 35 tools,
of which 9 respected their inclusion criteria
and were used to test several vulnerabilities on a sample set of 47,587 smart contracts.




\section{Conclusion}


To the best of our knowledge, this is the first work towards a static analysis framework for Tezos smart contracts.
\tezla positions itself as an intermediate representation obtained from a Michelson smart contract, the low-level language of Tezos smart contracts.
This representation abstracts the stack usage through the usage of a store, easing the adoption of mechanism and frameworks for program analysis that assume this characteristic, while maintaining the original semantics of the smart contract.

We have presented a case study on how this intermediate representation can be used to implement a static analysis by using \tezla along side the \softc platform.
This has shown how effortlessly one can perform static analysis on Michelson code without forcing developers to use a different language or implement \texttt{ad-hoc} static analysis tooling for a stack based language.

Michelson smart contracts have a mechanism of contract level polymorphism called entrypoints, where a contract can be called with an entrypoint name and an
argument.
This mechanism takes the form of a parameter composed as nesting of \texttt{or} types with entrypoint name annotations.
This parameter is then checked at the top of contract in a nesting of \texttt{IF_LEFT} instructions, running the desired entrypoint this way.
This mechanism is optional and transparent to smart contracts without entrypoints.
As such, they are also transparent to \tezla.
We therefore plan to extend \tezla in order to deal with entrypoints and generate isolated components for each entrypoint of a smart contract, which allow us to obtain clearer control-flow graphs and analysis
results.

Future plans include a formal account of the \tezla resource
analysis in order to formally verify that the semantics (including gas
consumption) of a \tezla-represented contract are maintained in respect to the
original Michelson code. This will also make way to the development of a
platform for principled static analysis of Michelson smart contracts.
We plan to study which properties are of interest so that we can integrate existing tools and algorithms for code optimization, resource usage analysis and security and correctness verification.

Another direction to tackle is the interfacing of \tezla with other static analysis platforms such as those provided by the MOPSA project \cite{mine-TAPAS18} which, among other abilities, provides a means to integrate static analyses.


\bibliographystyle{abbrvurl}
\bibliography{bibliography}%

\begin{thebibliography}{10}

\bibitem{fresco}
{{FRESCO}} - formal verification and static analysis of tezos compliant smart
  contracts, gitlab.
\newblock URL: \url{https://gitlab.com/releaselab/fresco}.

\bibitem{michelson}
Michelson: The language of {{Smart Contracts}} in {{Tezos}}.
\newblock URL: \url{http://tezos.gitlab.io/whitedoc/michelson.html}.

\bibitem{tezcheck}
{{TezCheck}} - softcheck interface for tezos, gitlab.
\newblock URL: \url{https://gitlab.com/releaselab/fresco/tezcheck}.

\bibitem{Alfour}
G.~Alfour.
\newblock {{LIGO}}.
\newblock URL: \url{https://ligolang.org/}.

\bibitem{Andrews}
S.~Andrews and R.~Ayotte.
\newblock Fi - {{Smart}} coding for smart contracts.
\newblock URL: \url{https://fi-code.com/}.

\bibitem{Antonino2020}
P.~Antonino and A.~W. Roscoe.
\newblock Formalising and verifying smart contracts with {{Solidifier}}: A
  bounded model checker for {{Solidity}}.
\newblock Feb. 2020.
\newblock \href {http://arxiv.org/abs/2002.02710} {\path{arXiv:2002.02710}}.

\bibitem{Bernardo2019}
B.~Bernardo, R.~Cauderlier, Z.~Hu, B.~Pesin, and J.~Tesson.
\newblock Mi-{{Cho}}-{{Coq}}, a framework for certifying {{Tezos Smart
  Contracts}}.
\newblock In {\em 1st {{Workshop}} on {{Formal Methods}} for {{Blockchains}}},
  Sept. 2019.
\newblock \href {http://arxiv.org/abs/1909.08671} {\path{arXiv:1909.08671}}.

\bibitem{Bernardo2020}
B.~Bernardo, R.~Cauderlier, B.~Pesin, and J.~Tesson.
\newblock Albert, an intermediate smart-contract language for the {{Tezos}}
  blockchain.
\newblock In {\em 4th {{Workshop}} on {{Trusted Smart Contracts}}}, Jan. 2020.
\newblock \href {http://arxiv.org/abs/2001.02630} {\path{arXiv:2001.02630}}.

\bibitem{DaiLambda}
DaiLambda.
\newblock {{SCaml}}.
\newblock URL: \url{https://gitlab.com/dailambda/scaml}.

\bibitem{Durieux2020}
T.~Durieux, J.~F. Ferreira, R.~Abreu, and P.~Cruz.
\newblock Empirical {{Review}} of {{Automated Analysis Tools}} on 47,587
  {{Ethereum Smart Contracts}}.
\newblock In {\em 42nd {{International Conference}} on {{Software Engineering}}
  ({{ICSE}} '20)}, Feb. 2020.
\newblock \href {http://arxiv.org/abs/1910.10601} {\path{arXiv:1910.10601}},
  \href {https://doi.org/10.1145/3377811.3380364}
  {\path{doi:10.1145/3377811.3380364}}.

\bibitem{Feist2019}
J.~Feist, G.~Greico, and A.~Groce.
\newblock Slither: A static analysis framework for smart contracts.
\newblock In {\em Proceedings of the 2nd {{International Workshop}} on
  {{Emerging Trends}} in {{Software Engineering}} for {{Blockchain}}},
  {{WETSEB}} '19, pages 8--15, {Montreal, Quebec, Canada}, May 2019. {IEEE
  Press}.
\newblock \href {https://doi.org/10.1109/wetseb.2019.00008}
  {\path{doi:10.1109/wetseb.2019.00008}}.

\bibitem{Goodman2014}
L.~M. Goodman.
\newblock Tezos-a self-amending crypto-ledger {{White}} paper.
\newblock Technical report, 2014.

\bibitem{Kam1977}
J.~B. Kam and J.~D. Ullman.
\newblock Monotone data flow analysis frameworks.
\newblock {\em Acta Informatica}, 7(3):305--317, 1977.
\newblock \href {https://doi.org/10.1007/BF00290339}
  {\path{doi:10.1007/BF00290339}}.

\bibitem{Maurel}
F.~Maurel and S.~C. Arena.
\newblock {{SmartPy}}.
\newblock URL: \url{https://smartpy.io/}.

\bibitem{mine-TAPAS18}
A.~Min{\'e}, A.~Ouadjaout, and M.~Journault.
\newblock Design of a modular platform for static analysis.
\newblock In {\em Proc{.}~of 9h Workshop on Tools for Automatic Program
  Analysis (TAPAS'18)}, Lecture Notes in Computer Science (LNCS), page~4, 28
  Aug{.} 2018.
\newblock \url{http://www-apr.lip6.fr/~mine/publi/mine-al-tapas18.pdf}.

\bibitem{reisSoftcheck}
J.~Reis.
\newblock Softcheck, a generic and modular data-flow analyser.
\newblock URL: \url{https://gitlab.com/joaosreis/softcheck}.

\bibitem{ReisTezCheckAnalysisResultsGitLab}
J.~Reis.
\newblock {{TezCheck Analysis Results}} - {{GitLab}}.
\newblock URL: \url{https://gitlab.com/releaselab/fresco/tezcheck-results}.

\bibitem{Rosen1988}
B.~K. Rosen, M.~N. Wegman, and F.~K. Zadeck.
\newblock Global value numbers and redundant computations.
\newblock In {\em Proceedings of the 15th {{ACM SIGPLAN}}-{{SIGACT}} Symposium
  on {{Principles}} of Programming Languages - {{POPL}} '88}, pages 12--27,
  {San Diego, California, United States}, 1988. {ACM Press}.
\newblock \href {https://doi.org/10.1145/73560.73562}
  {\path{doi:10.1145/73560.73562}}.

\bibitem{Sergey2018}
I.~Sergey, A.~Kumar, and A.~Hobor.
\newblock Scilla: A {{Smart Contract Intermediate}}-{{Level LAnguage}}.
\newblock 2018.
\newblock \href {http://arxiv.org/abs/1801.00687} {\path{arXiv:1801.00687}}.

\bibitem{sergeySaferSmartContract2019}
I.~Sergey, V.~Nagaraj, J.~Johannsen, A.~Kumar, A.~Trunov, and K.~C.~G. Hao.
\newblock Safer smart contract programming with {{Scilla}}.
\newblock {\em Proceedings of the ACM on Programming Languages},
  3(OOPSLA):1--30, Oct. 2019.
\newblock \href {https://doi.org/10.1145/3360611} {\path{doi:10.1145/3360611}}.

\bibitem{Serokell}
Serokell and T.~Group.
\newblock Lorentz.
\newblock URL:
  \url{https://gitlab.com/morley-framework/morley/-/tree/master/code/lorentz}.

\bibitem{Szabo1997}
N.~Szabo.
\newblock Formalizing and {{Securing Relationships}} on {{Public Networks}}.
\newblock {\em First Monday}, 2(9), Sept. 1997.
\newblock \href {https://doi.org/10.5210/fm.v2i9.548}
  {\path{doi:10.5210/fm.v2i9.548}}.

\end{thebibliography}
\end{document}